\documentclass[intlimits,twoside,a4paper]{article}

\usepackage{amsmath,amssymb}
\usepackage{graphicx}
\usepackage{wrapfig}

\usepackage[T2A]{fontenc}
\usepackage[cp1251]{inputenc}

\usepackage[eqsecnum]{cmpj2}

\issue{2012}{15}{4}{43604}

\doinumber{10.5488/CMP.15.43604}

\title[Towards frustration of freezing transition in a binary hard-disk mixture]%
{Towards frustration of freezing transition in a binary hard-disk mixture}
\author[A. Huerta, V. Carrasco-Fadanelli, A. Trokhymchuk]{A. Huerta\refaddr{label1},
        V. Carrasco-Fadanelli\refaddr{label1}, A. Trokhymchuk\refaddr{label2}}
\addresses{
\addr{label1} Facultad de F\'isica e Inteligencia Artificial,
Depto. de F\'isica, Universidad Veracruzana,\\
Circuito Gonz\'alo Aguirre Beltr\'an s/n Zona Universitaria
Xalapa, Ver. C.P. 91000, M\'exico.
\addr{label2} Institute for Condensed Matter Physics
of the National Academy of Sciences of Ukraine,\\
Department of the Theory of Solutions,
1 Svientsitskii St., 79011 Lviv, Ukraine
}

\date{Received September 27, 2012}
\authorcopyright{A. Huerta, V. Carrasco-Fadanelli, A. Trokhymchuk, 2012}

\begin{document}

\maketitle

\begin{abstract}

The freezing mechanism, recently suggested for a monodisperse hard-disk fluid
[Huerta et al, Phys. Rev. E, 2006, \textbf{74}, 061106] is extended here to an
equimolar binary hard-disk mixtures.
We are showing that for diameter ratios, smaller than 1.15 the global orientational order parameter
of the binary mixture behaves like in the case of a monodisperse fluid. Namely, by increasing the
disk number density there is a tendency to form a crystalline-like phase.
However, for diameter ratios larger than 1.15 the binary mixtures behave like a disordered fluid.
We use some of the structural and thermodynamic properties to compare and discuss the behavior as a function of diameter ratio and packing fraction.
\pacs 64.60.Fr, 68.35.Rh
\end{abstract}

\section{Introduction}

One-component or monodisperse system of hard spheres is the simplest and the most popular model system in condensed
matter physics.
The only feature that distinguishes a hard-sphere model from the ideal system is a non-zero hard-core diameter.
Because of this feature the hard-sphere model already exhibits a set of fundamental properties observed
for a real condensed matter such as liquid-like short-range ordering and liquid-to-solid transition.
A number of papers and textbooks can be found in the literature that are devoted to the discussion of the hard-sphere model
and its properties~\cite{BarkerHenderson,HansenMcdonalds}.

A much simpler and more transparent system seems to be a two-dimensional counterpart of the hard-sphere model, i.e.,
just one layer of hard spheres or, which in this case is the same, an assembly of hard disks spread out in
the plane.
In contrast to the ideal system and similarly to the hard-sphere model, by increasing the number of disks, this system
undergoes transformation from a disordered fluid into a perfect two-dimensional crystal. This transformation is accompanied
by a freezing transition which was  first evidenced by Alder and Wainwright~\cite{AlderWainwright} fifty years ago.
By means of the molecular dynamics simulations Alder and Wainwright have shown that
a system of hard disks has a density region where its pressure isotherm exhibits a van der Waals-like loop.
Since then, a number of papers have been published~\cite{Binder,Truskett1998,Huerta2006} discussing
various aspects of this phenomenon, such as the origin of the freezing transition,
the mechanisms of freezing, the criteria necessary for the freezing transition to occur, etc.
To date, there is a consensus that the freezing transition in a hard-disk system is driven by entropy,
and precise locations of the freezing and melting
densities are detected~\cite{Binder,Truskett1998,Huerta2006}.

\begin{figure}
\includegraphics[width=0.47\textwidth]{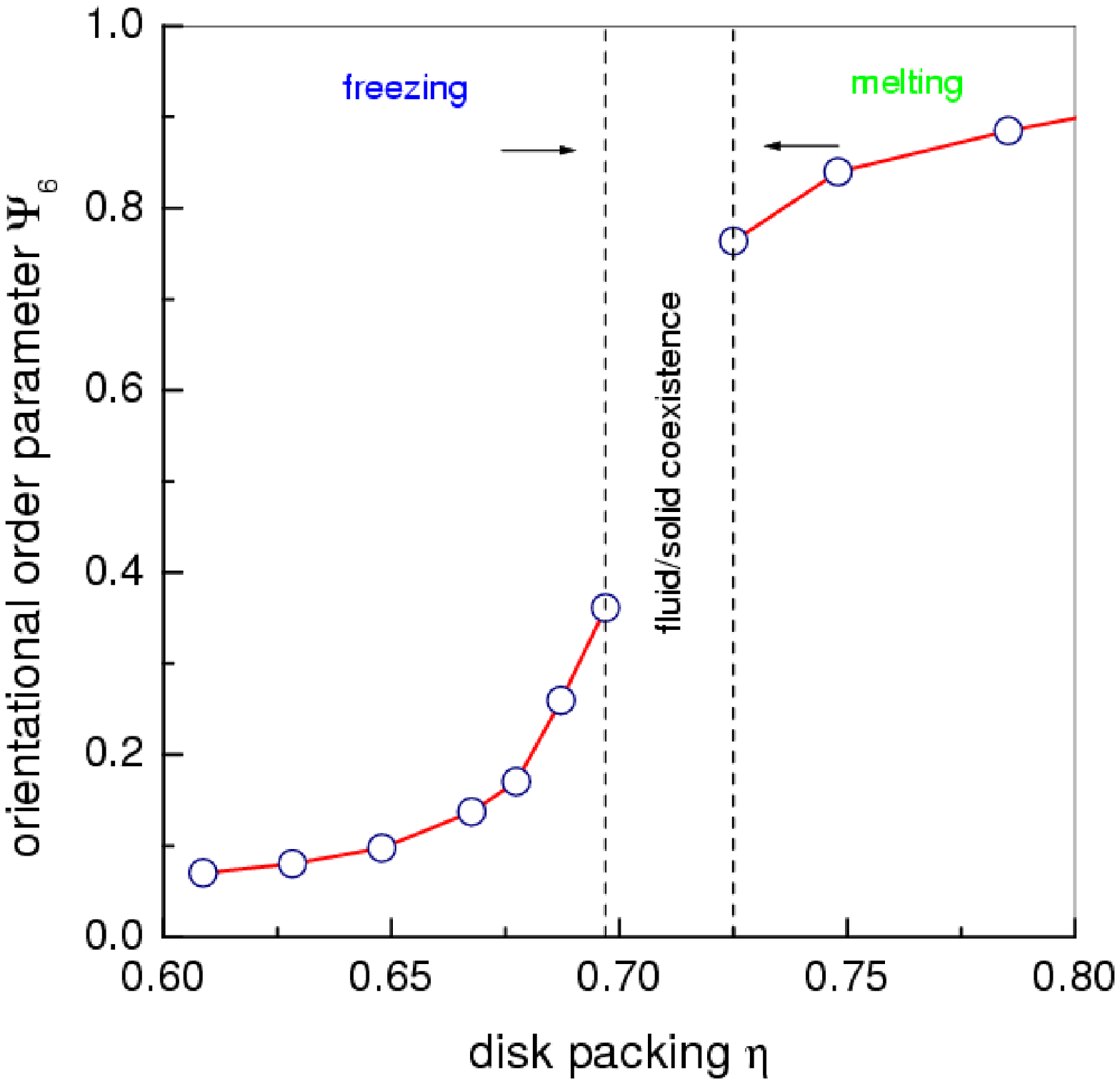}%
\hspace{1cm}%
\includegraphics[width=0.44\textwidth]{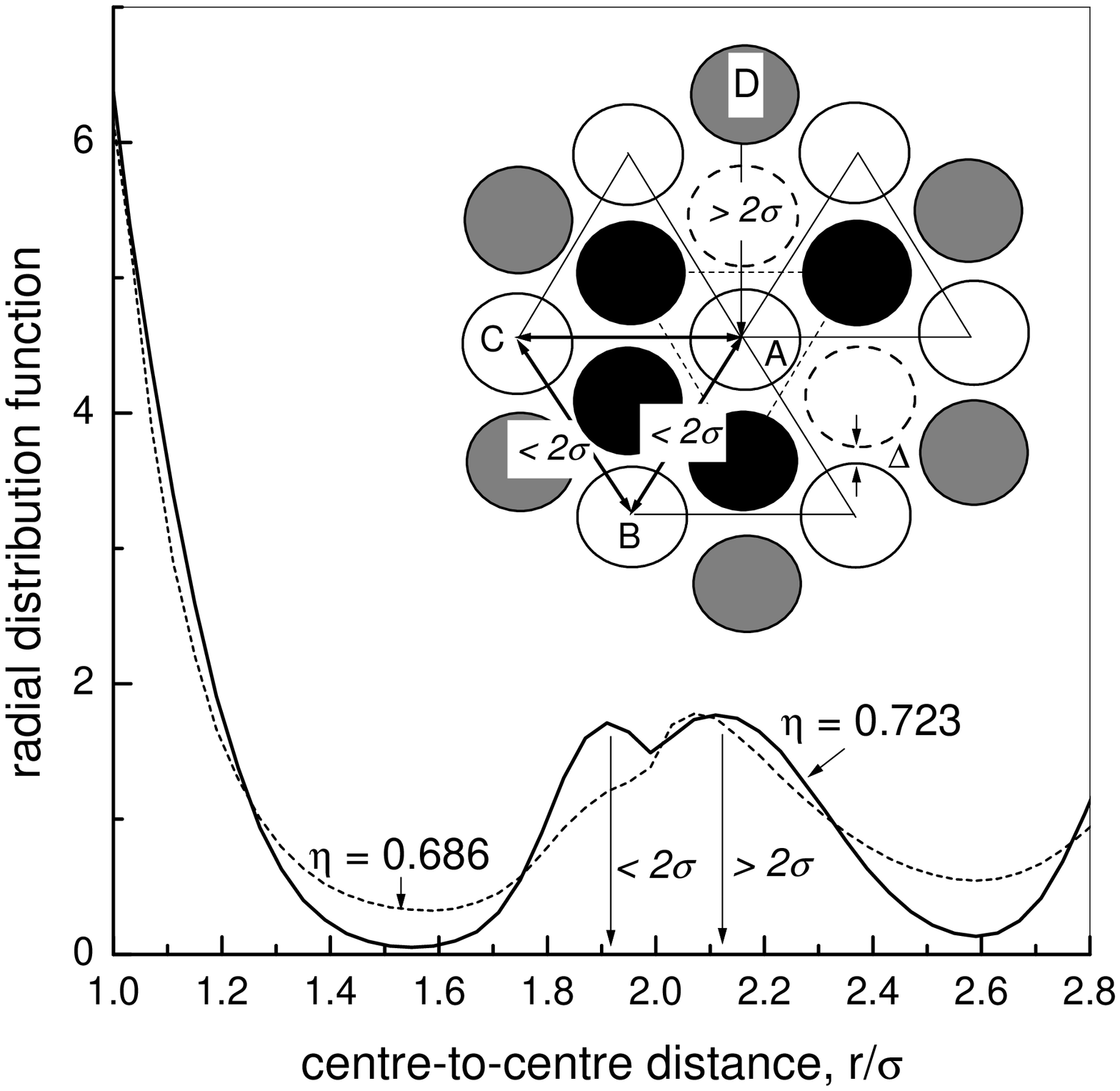}%
\hfill\\%
\parbox[t]{0.47\textwidth}{%
\centerline{(a)}%
}%
\hspace{1cm}%
\parbox[t]{0.44\textwidth}{%
\centerline{(b)}%
}\hfill%
\caption{Monodisperse disk system. Part a) Global bond-orientational order parameter $\psi_6$ as a function
of packing fraction $\eta$.
Part b) Disk-disk radial distribution function $g(r)$ at two packing fractions, namely, $\eta=0.686$ that is just before
freezing transition  and $\eta=0.723$ which is slightly higher than the melting density in this system~\cite{Huerta2006}.
The inset shows a schematic of the central disk (the hollow disk A at the center) and its first
(the disks filled with black color and hollow disks with dash line) neighbors
and the second (the disks filled with gray color and hollow disks with dashed line) neighbors,
or the first and the second coordination shells, respectively.
}  \label{RDF}
\end{figure}

Besides the pressure
calculations that were  originally employed by Alder and Wainwright~\cite{AlderWainwright},
another common way to illustrate the transition from a disordered
fluid to an ordered solid in two-dimensional monodisperse system is the density evolution of
the global-bond orientational order parameter~\cite{Binder,Kansal}.
The global orientational order parameter quantifies the degree of deviation of the nearest neighbors arrangement of each particle
in the system from the perfect hexagonal arrangement. The highest value of this parameter equals unity and corresponds to a
perfect hexagonal crystal.
The case of a monodisperse hard-disk system is illustrated in figure~\ref{RDF}~(a).
The formation of local quasiregular hexagonal ordering with increasing density in a hard disk system [see the inset in figure~\ref{RDF}~(b)]
implies that the
average center-to-center distance between any disk chosen as central and its closer second coordination shell neighbors
[the open, solid line disks in the inset in figure~\ref{RDF}~(a)] becomes shorter than two disk diameters.
This fact is consistent with the shape of the disk-disk radial distribution function, reflecting itself in the appearance
of the shoulder on its second maximum.

On the other hand, within the hexagonal arrangement, each central disk and its closer second coordination shell neighbors
 simultaneously serve as the alternating nearest neighbors of the common neighboring disk.
When the centre-to-centre distance between
alternating nearest neighbors becomes shorter than two disk diameters, this common
neighboring disk becomes caged.

Based on this observation, we have recently reported a simple mechanism for the freezing
of hard disks~\cite{Huerta2006}. This mechanism considers that by taking only alternating nearest neighbors
into account it is sufficient to describe the fluid and solid phases as well as the transition between them.
The fluid becomes unstable when the average centre-to-centre distance between
alternating nearest neighbors becomes shorter than two disk diameters and the resulting gap
between them is shorter than hard-core diameter and does not allow for the central disk to wander.
Such a caging concept allows for both the quantitative and qualitative description of the thermodynamics of
freezing transition in monodisperse hard-disk fluid and has been already utilized  to discuss percolation~\cite{Wang2011}.

We also note, that the appearance of a shoulder on the second peak of the radial distribution function
has been already suggested as a structural precursor to the freezing transition in
the hard disk system~\cite{Truskett1998} and is considered among others as empirical criteria for identification of
the freezing transition not only in the hard-disk system but more generally in the two-dimensional condensed matter~\cite{Wang2010}.


Such a detailed understanding of the freezing in a monodisperse hard-disk system implies that now one can try to move further and see how the freezing scenario that occurs here will be extended and/or modified going to more complex two-dimensional systems.
There are different possibilities to complicate the hard-disk model, e.g., by modifying the disk-disk interaction.
In this paper we will consider the simplest non-trivial complication of the monodisperse hard-disk system by introducing
the difference in the disk diameters.
Our goal is to investigate how the ordering behavior in a hard-disk system will change the transition from a monodisperse
hard-disk system to an equimolar binary mixture of hard disks of different diameters.

For a binary hard-disk mixture, it is already established that a progressive increase of the hard-core diameter for one half of the disks (equimolar mixture) does substantially change  the properties of this
system, especially, in the vicinity of the density region that corresponds to the freezing/melting transitions
in a monodisperse system.
In particular, following the results due to Speedy~\cite{Speedy1999}, the equimolar binary
hard-disk mixtures exhibit a freezing transition and proceed to a mixed crystal only for disk diameter ratios that do not exceed 1.2.
When the disk diameter ratio is exactly 1.2, the transition region shrinks, tending to disappear, while equilibrium freezing point was not precisely located in that study.
In the same paper~\cite{Speedy1999}, it is also shown that starting from the disk diameter ratio 1.3,
the equimolar binary mixtures pass into an amorphous state and it was suggested to consider them to be good glass-formers.

In what follows we will probe the behavior of the equimolar binary hard-disk mixture
by means of the global orientational order parameter and the disk-disk radial distribution functions.
In the next section~2, we describe the model and
outline the details of the computer simulation studies performed by us. The results and their discussion are presented in
section~3, while section~4 presents the summary and conclusions.



\section{Model definition and simulation details}

We consider a two-component (A and B) system of disks, at equimolar conditions, i.e.,
$\rho_A=\rho_B = \rho/2$, where $\rho$ is the total number density of disks.
The disks that belong to components A and B are of different diameters,
namely, $\sigma_A=\sigma+\delta$ and $\sigma_B=\sigma-\delta$.
The disk-disk pair interactions $U_{ij}(r)$, where $i$ and $j$ stand for the species $A$ and $B$, are given by
\begin{equation}
  U_{AA}(r) = \left\{
  \begin{array}{l l}
    \infty & \quad \text{if $r$ < $\sigma+\delta$}\,, \\
    0 & \quad \text{if $r$ > $\sigma+\delta$}\,,\\
  \end{array} \right.
\end{equation}
\begin{equation}
  U_{BB}(r) = \left\{
  \begin{array}{l l}
    \infty & \quad \text{if $r$ < $\sigma-\delta$}\,,\\
    0 & \quad \text{if $r$ > $\sigma-\delta$}\,,\\
  \end{array} \right.
\end{equation}
and
\begin{equation}
  U_{AB}(r)=U_{BA}(r) = \left\{
  \begin{array}{l l}
    \infty & \quad \text{if $r$ < $\sigma$}\,,\\
    0 & \quad \text{if $r$ > $\sigma$}\,,\\
  \end{array} \right.
\end{equation}
where $r$ is the center-to-center separation.
Useful quantities to characterise the system will be the total packing fraction
given by $\eta=(\pi/4)\rho(\sigma^2+\delta^2)$ and the disk diameter ratio $R=(\sigma+\delta)/(\sigma-\delta)$.
In what follows we will vary the parameter $\delta$ from $\delta=0$, which corresponds to a monodisperse system,
up to $\delta=0.166$ which corresponds to a binary disk mixture having size ratio $R=1.4$.
The parameter $\sigma$ will be used as a unit length throughout the paper.

Under certain pressure conditions the system of hard disks exhibits a fluid-to-solid transition.
To simulate such a system, a number $\,N\,$ of hard disks  were placed into a squared area of the size
$\,A=L\times L\,$  and periodic boundary conditions in both directions were applied.
A standard Monte Carlo simulation technique using Metropolis
algorithm was used to obtain ensemble averages of the dense
equilibrium hard disk system.

Two types of computer simulations have been carried out.
To evaluate the structural properties we have used an NVT ensemble
with 400 particles. Each equilibration run was relaxed for at least
$10^7$ iterations per particle. For productive runs, we have averaged
over at least 20000 different configurations, each being relaxed by 1000
iterations per particle. The acceptance ratio has been fixed between $20\%$ and $30\%$
and was controlled by choosing the maximum displacement of at least $13\%$.

To calculate the thermodynamical properties we have employed a NPT ensemble
which consists in the translation and change of the volume (area).
The Monte Carlo simulations in the NPT ensemble were aimed at localizing the fluid-solid transition region.
To obtain an ensemble averaging of each state, we have to proceed in the
following way. Starting with fluid densities at equilibrium, obtained
from the NVT ensemble, we fix a fluid pressure attempting to translate
each particle for at least 1000, followed by an attempt to change the
simulation box size. Translation and box size acceptance ratio were
controlled choosing the maximum displacements of particles and box
size, for at least $13\%$ and $30\%$, respectively. Averages were stored for
at least 30000 iterations.
We have  also verified that the attempts to interchange the particle
identity do not change the results at the higher packing fractions
studied.


\section{Results and discussion}

We explore the effect of the disk diameter ratio on the hexagonal ordering phenomenon in the equimolar binary hard-disk  mixture by
performing Monte Carlo simulation studies of a set of six
values of the size ratio parameter $\delta = 0.02,$ 0.05, 0.07, 0.08, 0.09 and 0.166
that correspond to the diameter ratios $R=1.04,$ 1.11, 1.15, 1.17, 1.2 and 1.4, respectively.
We restricted the present study to the maximal size ratio $R=1.4$, since it was already suggested by Speedy~\cite{Speedy1999}
that such a binary hard-disk mixture can be considered as a glassforming fluid.
While analyzing the obtained results we found that the equimolar binary hard-disk mixtures having size ratios that do not
exceed $R=1.15$, behave qualitatively similarly to a monodisperse hard-disk system. Thus, in what follows
we will discard
from the discussion the results for the diameter size ratios $R=1.04$ by presenting only the
results for the case $R=1.11$ as being  typical of this class of the equimolar binary hard-disk mixtures.
Similarly, we will use mainly the case of $R=1.2$ to discus the mixtures having diameter size ratios that
exceed the value $R=1.15$.

\begin{figure}[!b]
\centerline{\includegraphics[width=0.6\textwidth]{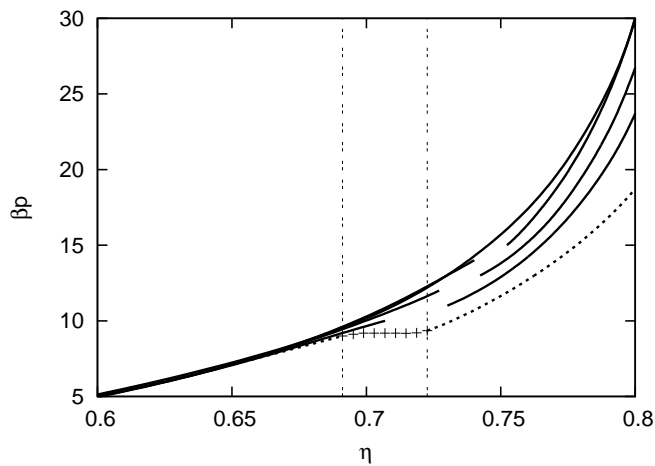}}
\caption{Equation of state of the equimolar binary hard-disk mixtures with diameter
size ratios $R=1.4$, $1.2$, $1.15$, $1.11$ (from the top to the bottom) in comparison
against a monodisperse, $R=1$, hard-disk system (the dotted line from the bottom).
The two thin vertical lines, from left to right, mark the packing fractions,
$\eta_{\rm f}=0.69$ and $\eta_{\rm m}=0.723$, that correspond to the
freezing and melting points, respectively, in the case of a monodisperse hard-disk system;
the density region between these two lines is the fluid-solid transition region and the results shown by symbols
are taken from the extensive computer simulation study due to Mak~\cite{Mak}.
For all other systems the fluid-solid transition regions correspond to the empty space between fluid and solid
pressure curves. In the case of the mixture with diameter size ratio $R=1.4$ transition region was not detected.
} \label{termo}
\end{figure}

\subsection{Equation of state}

To present our results, we start with the Monte Carlo simulation data obtained for the equation of state of
the hard-disk systems under consideration.
One of the purposes of the pressure calculations was to make a reference to the work by
Speedy~\cite{Speedy1999} who calculated the equation of state for hard-disk mixtures by means
of molecular dynamics simulations.
It was suggested~\cite{Speedy1999} that a freezing transition, observed for a monodisperse hard-disk system,
can be frustrated in a binary hard-disk mixture by increasing the ratio of disk diameters.
The highest disk size asymmetry that still
allows for a freezing transition in a binary hard-disk mixture to occur was found to be 1.2:1,
while for the size ratio 1.4:1 the
glassy state can be formed by detouring the freezing transition.


Figure~\ref{termo} shows the pressure versus density data obtained
for four equimolar binary hard-disk mixtures having diameter ratios $R=1.11,$ 1.15, 1.2 and 1.4.
For comparison and for better understanding,
the corresponding results for the case of a monodisperse hard-disk system, $R=1$, taken from our earlier
study~\cite{Huerta2006}, are shown in figure~\ref{termo} as well.
The two thin vertical lines in this figure, from left to right, mark the packing fractions,
$\eta_{\rm f}=0.69$ and $\eta_{\rm m}=0.723$, that correspond to the
freezing and melting points, respectively, in the case of a monodisperse hard-disk system.
The density region between these two lines refers to the fluid-solid transition region for hard disks of the
same diameters.
From the results presented in figure~\ref{termo} it follows that the disk diameter asymmetry does not affect
the pressure for the densities that precede this transition region.
 Namely, at $\eta < \eta_{\rm f}$,  all hard-disk systems, considered in the present study, exhibit nearly the same
pressure profile, showing smooth increment with an increase of density.
It is only at densities $\eta > \eta_{\rm f}$ that notable differences in the pressure caused by disk size asymmetry are observed.

To locate the possible freezing and melting  points and to determine the transition region
in the case of a hard-disk mixture, we carried out a series of constant pressure (NPT)
Monte Carlo simulation runs for each of the binary systems under study.
By smoothly incrementing a fixed pressure value, we were trying
to find the pressure $P$ at which the disk density, that  was also changing smoothly, makes a jump
to a notably higher value. These values of the pressure and density might be referred to as the upper bounds
for the  pressure and density at melting point, $P_{\rm m}$ and $\eta_{\rm m}$, respectively.
The pressure $P$ that preceded such a density jump, and corresponding to this pressure density, might be
referred to as the bottom bounds for the  pressure and density at
freezing point, $P_{\rm f}$ and $\eta_{\rm f}$, respectively. By proceeding in this way, we were able to
locate the transition regions in all binary hard-disk mixtures considered, except the one that corresponds
to the disk diameter ratio $R=1.4$.  In figure~\ref{termo}, these transition regions can be seen as an empty space between the corresponding fluid pressure and solid pressure curves.
A complete  set of data, that include freezing and melting densities and pressures for all systems
considered in this study, are collected in table~\ref{tab}.

%
\begin{table}[tb]
\centering{
\caption{Pressure and packing fraction at the freezing and melting points of
the equimolar binary hard-disk mixtures with
different ratios of diameters. The data for a monodisperse disk system ($R=1$)
are taken from~\cite{Huerta2006}
and  are given for comparison.
\label{tab}}
\vspace{2ex}
\begin{tabular}{|ccccccccccccc|c}
\hline 
    $\frac{}{}R=\sigma_{\rm A}/\sigma_{\rm B}$ & & 1 & & 1.04 & & 1.11 & & 1.15 & & 1.17 & & 1.2 \\
\hline \hline 
 \multicolumn{1}{|c}{Freezing} &&&&&&&&&&&&\\
    $P_{\rm }/kT$ && 9.0 && 9.0 && 10.0 && 12.0 && 13.0 && 14.0 \\
    $\eta_{\rm }$ && 0.692 && 0.692 && 0.707 && 0.727 && 0.735 && 0.740 \\[2ex]
\hline
  \multicolumn{1}{|c}{Melting}&&&&&&&&&&&& \\
    $P_{\rm }/kT$ && 10.0 && 10.0 && 11.0 && 13.0 && 14.0 && 15.0 \\
    $\eta_{\rm }$ && 0.724 && 0.724 && 0.730 && 0.742 && 0.747 && 0.752 \\
\hline
\end{tabular}
}
\end{table}

From the results presented in table~\ref{tab} one can see that both the freezing and the melting point densities increase with an increase of a disk size asymmetry. Consequently, the transition regions
(see figure~\ref{termo} for details) are also shifted
to higher densities, with clear tendency to shrink and, finally, to disappear in the case of the
highest disk size asymmetry considered in the present study, i.e., at $R=1.4$.
We note that pressure calculations for the same hard-disk systems were also performed from molecular
dynamics simulations.
This has been done by taking the final disk configurations generated during the NPT Monte Carlo runs.
The results obtained from both techniques are identical within the margin of error.




\subsection{Orientational ordering}

Following the case of a monodisperse disk system shown in figure~\ref{RDF}, the freezing transition in a binary hard-disk
mixture can be probed by mapping the density dependence of the orientational ordering of the disks.
A commonly used quantitative measure of the orientational order in two-dimensional systems is the global
bond-orientational order parameter~\cite{Binder,Kansal}.
This parameter was evaluated during the Monte Carlo simulation runs using the following definition,
\begin{equation}
\psi_6=\left|\frac{1}{N_{nn}}\sum_{j}\sum_{k}\re^{6i\theta_{jk}}\right|,
\end{equation}
where $j$ runs over all disks in the system, $k$ runs over all ``geometric'' nearest
neighbours (nn) of disk $j$, each obtained through the Voronoi analysis, and $N_{nn}$ is
the total number of such nearest neighbours in the system.
The angle $\theta_{jk}$ is defined between some fixed reference axis in the system and
the vectors (``bonds'') connecting the nearest neighbours $j$ and $k$.

\begin{figure}[!h]
\centerline{\includegraphics[width=0.6\textwidth]{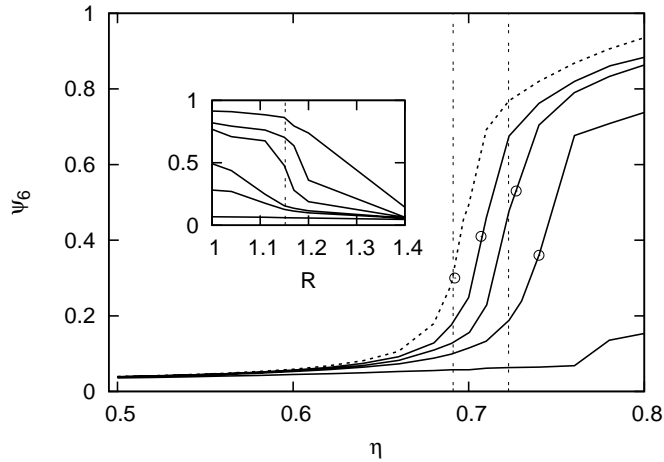}}
\caption{Global bond-orientational order parameter $\psi_6$
of equimolar binary hard-disk mixtures having diameter
size ratios $R=1.4, \ 1.2, \ 1.15, \ 1.11$ (from the bottom to the top) in comparison
with a monodisperse, $R=1$, hard-disk system (the dotted line from the top).
The two thin vertical lines mark the same as in figure~\ref{termo}, while open circles mark
the magnitude of $\psi_6$ at freezing packing fraction for a corresponding system.
The inset shows $\psi_6$ as a function of the disk diameter ratio $R$ for the selected packing fractions
from the bottom to the top $\eta=0.62, \ 0.69, \ 0.70, \ 0.723, \ 0.74$ and $0.80$.
 }  \label{PSI6}
\end{figure}

In figure~\ref{PSI6} we show the density dependence of the global bond-orientational
order parameter $\psi_6$ for binary hard-disk mixtures having diameter size ratios $R=1.11,$ 1.15, 1.2 and 1.4.
The corresponding results for a monodisperse hard-disk systems, $R=1$, are presented in figure~\ref{PSI6} as well.
As expected, at low densities and independently of the disk diameter ratio, all hard-disk systems behave
like a two-dimensional isotropic fluid. As the density of disks increases, both the monodisperse hard-disk system
and the binary hard-disk mixture  tend to form a hexagonal ordering.
However, in the case of a binary mixture having the disk diameter ratio $R=1.4$, this tendency is notably weaker, including
the highest density, $\eta=0.8$, that was probed in this study.

Overall, by comparing the results for $\psi_{6}$ in the case of binary hard-disk mixtures
at the same packing fraction with the results for the monodisperse disk system
 we can conclude that an increase of the disk diameter asymmetry promotes a decrease of the
hexagonal ordering in the system. However, the rate of this decrease depends on the disk diameter ratio.
At the same time, determining the magnitude of the hexagonal ordering in hard-disk systems at
the freezing point (open circles in figure~\ref{PSI6}) one can see that  $\psi_{6}$ initially increases with
an increase of the disk size asymmetry.
However, for the diameter size ratios larger than $R=1.15$, the hexagonal ordering at the freezing point starts to decline.

In the inset of figure~\ref{PSI6}, the global orientational order parameter is plotted
against the disk diameter ratio $R$ at several fixed packing fractions. One can see that at small
disk diameter asymmetry, the rate of a decline of the hexagonal ordering is very slow, but the values of $\psi_{6}$
significantly drop as the disk diameter ratio $R$ increases.
We notice that similarly to the results for the pressure in figure~\ref{termo}, the crossover between two different
behaviors of the global bond-orientational order parameter $\psi_{6}$ takes place
around the disk diameter ratio $R=1.15$.

\subsection{Radial distribution functions}

The fact that pressure isotherms for hard-disk mixtures in figure~\ref{termo} is quite similar to that
of the monodisperse disk system at low packing fractions, but
drastically changes when the disk packing fraction $\,\eta\,$ exceeds the value $\eta_{\rm f}=0.692$
at freezing point of a monodisperse disk system, may indicate that the freezing mechanism that
was recently suggested~\cite{Huerta2006} for a monodisperse hard-disk system undergoes important changes caused
by the disk size asymmetry.
The principal ingredient of the mechanism of a monodisperse hard disk freezing is the caging phenomenon
that occurs in the system of hard disks of equal diameters as a result of the entropy driven
hexagonal ordering. As it was already mentioned in the Introduction, this mechanism is quite
understandable and can be explained by means of the radial distribution
function of a monodisperse hard-disk system in the way shown in figure~\ref{RDF}~(b).

\begin{figure}[t]
\centerline{\bf R=1.11 \hspace{3.5cm} R=1.2 \hspace{3.5cm} R=1.4}
\centerline{\includegraphics[width=0.34\textwidth]{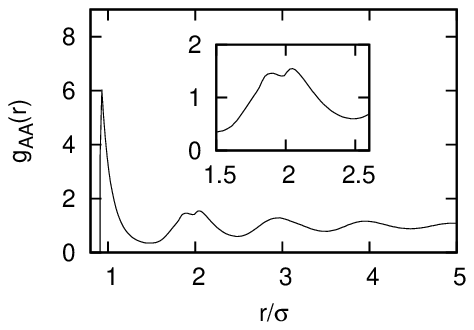}
\includegraphics[width=0.34\textwidth]{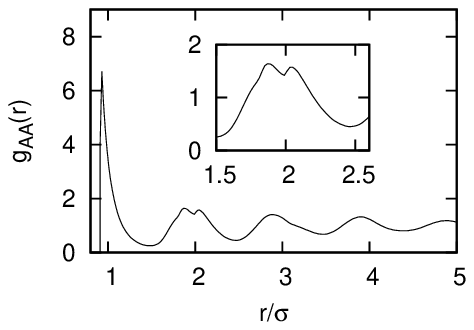}
\includegraphics[width=0.34\textwidth]{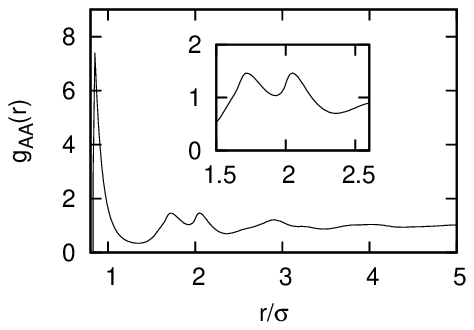}}
\centerline{\includegraphics[width=0.34\textwidth]{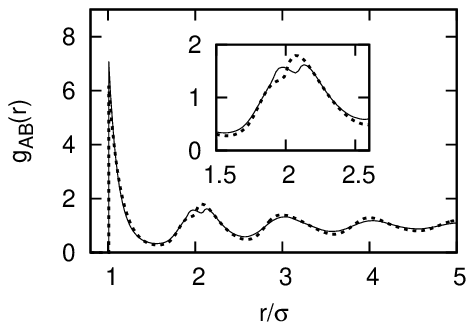}
\includegraphics[width=0.34\textwidth]{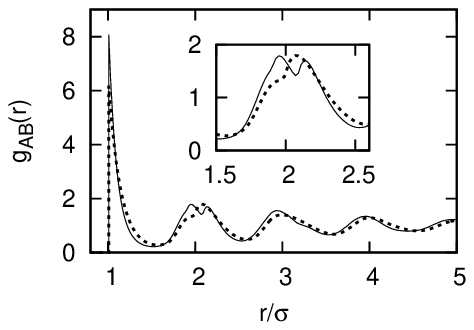}
\includegraphics[width=0.34\textwidth]{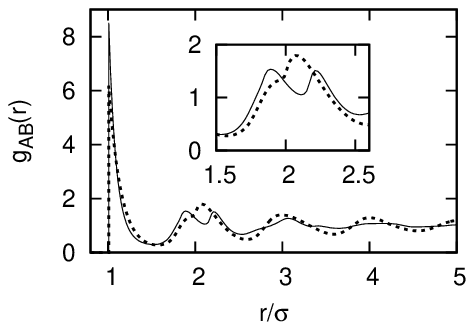}}
\centerline{\includegraphics[width=0.34\textwidth]{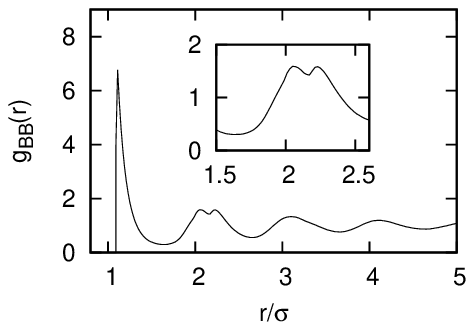}
\includegraphics[width=0.34\textwidth]{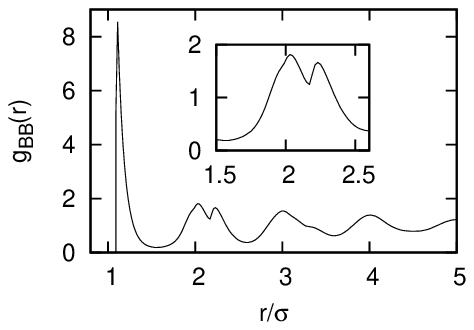}
\includegraphics[width=0.34\textwidth]{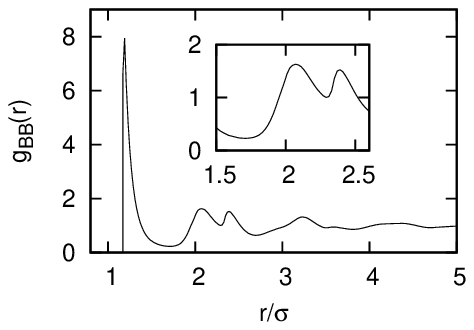}}
\caption{Radial distribution functions $g_{ij}(r)$
of equimolar binary hard-disk mixtures having disk diameter ratios $R=1.11, \ 1.2$ and $1.4$.
The dotted lines correspond to the radial distribution function $g(r)$ of the monodisperse hard-disk system at
$\eta= 0.692$ that is the freezing
point packing fraction of a monodisperse disk system,  while thin solid lines correspond to
the packing fractions at the
freezing point of particular binary hard-disk mixture, namely:
$\eta=0.702$ in the case of $R=1.11$, $\eta=0.740$ in the case of $R=1.2$; in the case of $R=1.4$, there is no
freezing point and we show the results for $\eta=0.740$.
All the insets show in detail the region of the second peak of the corresponding radial distribution function.
 } \label{grsall}
\end{figure}

To shed some light on the structural rearrangements that take place in binary hard-disk mixtures,
in figure~\ref{grsall} we present a set of radial distribution functions, $g_{AA}(r)$, $g_{BB}(r)$, and $g_{AB}(r)$
for three equimolar binary hard-disk mixtures characterized by the disk diameter ratios, $R=1.11,$ 1.2 and 1.4.
The first one, $R=1.11$, is a representative of the type of binary hard-disk mixtures for which the
freezing  and post-freezing features are similar to those of a monodisperse hard-disk system.
On the other hand, the second hard-disk system with $R=1.2$ represents the
type of binary hard-disk mixtures that still exhibit a freezing transition, but their properties
seem to be different from those of monodisperse hard-disk systems~\cite{footnote1}.
Finally, the third disk diameter ratio, $R=1.4$, corresponds to the binary hard-disk mixture that
does not experience a freezing transition, representing a glass-forming system~\cite{Speedy1999}.

The radial distribution functions shown in figure~\ref{grsall} are calculated at the
packing fraction values
that correspond to  the freezing point packing fraction
for each hard-disk mixture listed in table~\ref{tab}. Namely,
$\eta=0.707$ in the case of hard-disk mixture having the disk diameter ratio $R=1.11$,
and $\eta=0.74$ in the case of $R=1.2$.
Since there is no freezing transition in the case of a hard-disk mixture having the
disk diameter ratio $R=1.4$, the radial distribution functions for this system are obtained at
a packing fraction $\eta=0.74$.
  To make a reference to the case of all disks being of the same diameter, the radial distribution function $g_{AB}(r)$ for
each binary hard-disk system is explicitly compared with the radial distribution function $g(r)$ of the monodisperse
hard-disk system at the corresponding freezing point packing fraction, $\eta_{\rm f}=0.692$, for a monodisperse system.

First we note that the disk size asymmetry being introduced results in an increase of the values of the radial distribution
functions  $g_{ij}(r)$ at contact distances, $r=\sigma_{ij}$.
However, since the so-called contact value uniquely determines the pressure $\,P\,$ in
both the monodisperse and in the binary hard-disk systems,
from the results presented in figure~\ref{grsall} it follows that binary hard-disk mixtures require higher pressures
for the freezing transition to occur.
This is consistent with the results already discussed for the equation of state in figure~\ref{termo}.

More insight into the local structural rearrangement that presumably takes place in a
binary hard-disk mixture in the density region close to the freezing transition
can be achieved by analyzing the shape of the second maximum of the radial distribution functions.
It is quite evident that radial distribution functions $\,g_{\rm ij}(r)\,$
for each of three hard-disk mixtures show
a quite different and specific behavior, which is also different from
the radial distribution functions $\,g(r)\,$ of the monodisperse hard-disk system.
Nevertheless, we noticed that for a binary mixture having the disk diameter ratio $R=1.11$
(the first column in figure~\ref{grsall}),
 all three radial distribution functions , $g_{AA}(r)$, $g_{BB}(r)$ and $g_{AB}(r)$,
exhibit nearly the same shape of the second maximum.
Moreover, this shape obeys the shoulder, i.e., the feature that is discussed
in figure~\ref{RDF}~(b) and is responsible for the occurrence of the freezing transition in a monodisperse hard-disk
system~\cite{Huerta2006}.
However, speaking more generally we note that this suggestion by Truskett et al~\cite{Truskett1998}
towards the shoulder in the second maximum of the radial distribution function as the structural precursor of the
freezing transition does not seem to be general. In particular, it
could not be extended to the binary hard-disk mixtures with the disk diameter ratios $R > 1.2$, where both the shoulder
and splitting of the second maximum are extremely pronounced but freezing transition does not occur.

\section{Conclusions}

A binary equimolar mixture of hard disks having diameter ratios $R=1.04,$ 1.11, 1.15, 1.17, 1.2 and 1.4 have
been considered by means of computer simulations. The main purpose of this study was to probe these binary
mixtures on the freezing transition which normally occurs in a system of hard disks with the same diameters.
By calculating the equation of state we found that the size asymmetry being introduced alters the pressure in the systems.
We showed that a freezing transition shifts to higher disk packing fractions as the diameter asymmetry  increases
while the width of transition region shrinks.
The largest diameter ratio is $\,R=1.2\,$ when the freezing transition is still localized in hard-disk mixtures, while
the binary hard-disk mixture having diameter ratio $\,R=1.4\,$ does not exhibit the freezing behavior.
We consider that the change of the behavior towards the frustration of hexagonal order
is caused by the change of the maximum number of neighbors that each particle can have.
This is similar to the crossover observed earlier by introducing the short-ranged square-well attraction~\cite{Huerta2004} for the disk-disk interaction.


\section*{Acknowledgement}

This work is supported by the CONACYT of Mexico under the project 152431 and through the Red Tem\'atica de la Materia Condensada Blanda.


\begin{thebibliography}{99}

\bibitem{BarkerHenderson} Barker J.A., Henderson D., Rev. Mod. Phys., 1976, \textbf{48}, 587; \doi{10.1103/RevModPhys.48.587}.

\bibitem{HansenMcdonalds} Hansen J.-P., MacDonalds I.R., Theory of Simple Liquids, 3rd ed., Academic Press, 2006.

\bibitem{AlderWainwright} Alder B.J., Wainright T.E., Phys. Rev., 1962, \textbf{127}, 359; \doi{10.1103/PhysRev.127.359}.

\bibitem{Binder} Weber H., Marx D., Binder K., Phys. Rev. B, 1995, \textbf{61}, 14636; \doi{10.1103/PhysRevB.51.14636}.

\bibitem{Truskett1998} Truskett~T.M., Torquato~S., Sastry~S., Debenedetti~P.G., Stillinger~F., Phys. Rev. E, 1998, \textbf{58}, 3083; \\ \doi{10.1103/PhysRevE.58.3083}.

\bibitem{Huerta2006} Huerta~A., Henderson~D., Trokhymchuk~A., Phys. Rev. E, 2006, \textbf{74}, 061106;
\doi{10.1103/PhysRevE.74.061106}.

\bibitem{Kansal} Kansal~A.R., Truskett~T.M., Torquato~S., J. Chem. Phys., 2000, \textbf{113}, 4844; \doi{10.1063/1.1289238}.

\bibitem{Wang2011} Wang Z., Qi W., Peng Y., Alsayed A.M., Chen Y., Tong P., Han Y., J. Chem. Phys., 2011, \textbf{134}, 034506; \\ \doi{10.1063/1.3545967}.

\bibitem{Wang2010} Wang Z., Alsayed A.M., Yodh A.G., Han Y., J. Chem. Phys., 2010, \textbf{132}, 154501; \doi{10.1063/1.3372618}.

\bibitem{Speedy1999} Speedy~R.J., J. Chem. Phys., 1999, \textbf{110}, 4559; \doi{10.1063/1.478337}.

\bibitem{Mak} Mak C.H., Phys. Rev. E, 2006, \textbf{73}, 065104(R); \doi{10.1103/PhysRevE.73.065104}.

\bibitem{footnote1} The pressure isotherm in figure~\ref{termo} is approaching the pressure
isotherm of a glass-forming hard-disk mixture characterized by disk diameter ratio $R=1.4$.



\bibitem{Huerta2004} Huerta A., Naumis G.G., Wasan D., Henderson D., Trokhymchuk A., J. Chem. Phys., 2004, \textbf{120}, 1506; \\ \doi{10.1063/1.1632893}.





\end{thebibliography}

\ukrainianpart

\title{Про зникнення фазового переходу замерзання у бінарній суміші твердих дисків}
\author{А. Уерта\refaddr{label1}, В. Караско-Фаданеї\refaddr{label1}, А. Трохимчук\refaddr{label2}}
\addresses{
\addr{label1}Університет Веракрузана, Факультет фізики та інженерії, Кафедра фізики, \\
Халапа, Веракруз, СР 91000, Мексика
\addr{label2} Інститут фізики конденсованих систем НАН України,
Відділ теорії розчинів, \\ вул. Свєнціцького,~1,  79011~Львів, Україна
}

\makeukrtitle

\begin{abstract}
\tolerance=3000%
Механізм замерзання, який був запропонований недавно для однокомпонентного флюїду твердих дисків  [Huerta et al, Phys. Rev. E, \textbf{74}, 2006, 061106], узагальнено на випадок еквімолярної   бінарної суміші твердих дисків. Ми показуємо, що у випадку, коли відношення діаметрів є меншим за 1.15, то поведінка параметра глобального орієнтаційного порядку бінарної суміші є подібною до випадку однокомпонентного флюїду. А саме, при збільшенні густини дисків   спостерігається тенденція до утворення кристалоподібної фази. Однак, для відношень діаметрів більших ніж 1.15 відмічається зміна цієї тенденції на поведінку, характерну для невпорядкованого флюїду. Ми використовуємо окремі структурні та термодинамічні властивості для порівняння та обговорення їх поведінки як функції відношення діаметрів та параметра упаковки.
\keywords флюїд твердих дисків, переходи замерзання, еквімолярна бінарна суміш твердих дисків

\end{abstract}

\end{document}